\documentclass[manuscript,screen,nonacm]{acmart}
\AtBeginDocument{}

\setcopyright{acmlicensed}
\copyrightyear{2018}
\acmYear{2018}
\acmDOI{XXXXXXX.XXXXXXX}

\acmISBN{978-1-4503-XXXX-X/2018/06}

\usepackage{booktabs}
\usepackage{tabularx}
\usepackage{array}
\usepackage{ragged2e}

\begin{document}

\title[Click, Branch, Audit]{Click, Branch, Audit: A Decision-Tree Toolkit for Assessing Fundamental Rights Impacts under the Digital Services Act of Very Large Platforms \& Search Engines}
\titlenote{Preprint. Not peer reviewed.}

\author{Marie-Therese Sekwenz}
\affiliation{\department{Faculty of Technology, Policy and Management}
  \institution{Delft University of Technology}
  \city{Delft}
  \country{Netherlands}
}

\author{Till Winkler}
\affiliation{\institution{Interdisciplinary Transformation University Austria}
  \city{Linz}
  \country{Austria}
}

\author{Maria-Lucia Rebrean}
\affiliation{\institution{eLaw Center for Law and Digital Technologies, Leiden University}
  \city{Leiden}
  \country{Netherlands}
}

\author{Nina Baranowska}
\affiliation{\institution{Leiden University}
  \city{Leiden}
  \country{Netherlands}
}

\author{Manon Carrere}
\affiliation{\institution{Leiden University}
  \city{Leiden}
  \country{Netherlands}
}

\author{Gianclaudio Malgieri}
\affiliation{\institution{eLaw Center for Law and Digital Technologies, Leiden University}
  \city{Leiden}
  \country{Netherlands}
}

\author{Ben Wagner}
\affiliation{\institution{TU Delft}
  \city{Delft}
  \country{Netherlands}
}
\affiliation{\institution{Inholland}
  \city{Den Haag}
  \country{Netherlands}
}

\authorsaddresses{}
\renewcommand{\shortauthors}{Sekwenz et al.}

\begin{abstract}

The EU Digital Services Act (DSA) requires Very Large Online Platforms and Search Engines (VLOPs/VLOSEs) to assess and mitigate systemic risks, including actual or foreseeable negative effects on fundamental rights, and subjects these assessments to independent audit. Yet how such rights-relevant risks should be identified, assessed, evidenced, and documented remains methodologically under-specified. We address this gap by introducing a decision-tree-based toolkit that operationalizes Article 34(1)(b) DSA and related audit requirements into a structured assessment process. The toolkit decomposes platforms as systems-of-systems and guides assessors through \emph{DSA applicability, context definition, subsystem and stakeholder identification, risk identification, risk assessment, reporting,} and \emph{visualization}. A central design contribution is its treatment of vulnerability as situational and relational: rather than assigning vulnerability to fixed groups, the toolkit asks how particular platform subsystems and dependencies may place stakeholders in vulnerable positions. It further makes fundamental rights-relevant judgments traceable by recording the affected right, interference and justification analysis within the structure of a proportionality test, while also covering a \emph{perception} and \emph{impact} dimension including severity, scope, duration, reversibility, as well as practical consequences. We developed the toolkit iteratively through a design-science process combining literature-derived design requirements, three interdisciplinary expert workshops (N=4; N=11; N=20) testing the framework design with two DSA user persona scenarios. 
We contribute an extensible methodological foundation for making fundamental-rights risk assessments under the DSA more streamlined, systematic, transparent, and auditable.

\end{abstract}



\received{\today}

\maketitle

\section{Introduction}
Very Large Online Platforms (VLOPs) and Very Large Search Engines (VLOSEs) such as Meta, X, TikTok, Google, Amazon and others \cite{european_commission_overview_nodate} have become quasi-governmental actors. VLOPs not only wield enormous economic power, but also heavily influence fundamental rights~\cite{belli2022structural}. 
The European Union (EU) has adopted the Digital Services Act (DSA)~\cite{RegulationEU20222022} with the aim of creating a safer digital spaces in which fundamental rights are protected.\footnote{See Art.~1(1)~DSA: ``[...] setting out harmonised rules for a safe, predictable and trusted online environment that facilitates innovation and in which fundamental rights enshrined in the Charter, including the principle of consumer protection, are effectively protected."} 
The DSA  obliges VLOPs and VLOSEs with more than 45~million average active users in the EU (DSA~Art.~33~\cite{RegulationEU20222022})) to follow a set of refined obligations. 
These platforms have become central infrastructures that facilitate, communication with friends and family~\cite{auxier2021social}, trade~\cite{doi:10.1177/13548565231192103}, information retrieval~\cite{ConsumptionOnlineNews} and provide the infrastructure for political~\cite{kim2025mobilization} and civic participation in Europe~\cite{karakiza2015impact}. 
The EU has positioned itself as a regulatory pioneer in explicitly requiring providers to assess and mitigate systemic risks~\cite{griffin2025governing}, including negative impacts on the exercise of fundamental rights~\cite{micova2021harm}. Yet, translating these obligations into evaluative practice remains methodologically under-specified, particularly when it comes to capturing platform-driven, foreseeable 'negative effects' and risks as experienced by recipients of the service. 
Therefore a heart piece and key obligation of the DSA requires VLOPs/VLOSEs to identify, assess and mitigate systemic risks (DSA~Arts. 34-35~\cite{RegulationEU20222022}), including those with actual or foreseeable negative impacts on the exercise of fundamental rights (DSA~Art.~34(1)(b) ~\cite{RegulationEU20222022}). According to Art.~37 DSA, these systemic risk assessments must be evaluated through independent audits led by auditing firms as key practitioners for externally assessing compliance.
Acquiring expertise covering complex multi-actor auditing can be particularly difficult when the legal framework, such as the DSA, which is a novel regulation~\footnote{The DSA is fully in force since February 2024.} and therefore yields legal uncertainty and is lacking case law. 
Failure to compliance with the DSA can yield high fines (up to 6\% of annual global turnover), i.e. refrain from deceptive design (Art.~25(1)~DSA)—misleading ‘verified account’ signals—and shortcomings in the advertising repository and researcher data access (Arts.~39 and~40(12)~DSA) have already led to a €120~million fine for X (formerly Twitter)~\cite{CommissionFinesEU120}.
Existing work has not yet sufficiently operationalized how actual or foreseeable fundamental rights impacts, particularly for stakeholders (such as users) in vulnerable positions, can be systematically identified, evidenced, and assessed in socio-technical evaluations (like DSA systemic risk assessments and independent audits) beyond a primary focus on user-introduced harms for which
clear frameworks, standards, and practical toolkits remain limited in the context of new regulations like the DSA.  This toolkit and its implementation as a first proposition of an assessment framework with a step-by-step-workflow enables a shared understanding inviting interdisciplinary debate, iteration, and refinement of compliance processes that address the obligations of the Delegated Regulation (DR) for conducting audits~\cite{noauthor_delegated_2023}. 
This paper answers the core research question: 
\emph{
How can a structured risk-assessment methodology, (operationalized as a decision-tree-based toolkit), improve the identification, assessment, and documentation of adverse effects on fundamental rights, (under Art.~34(1)(b)~DSA) for VLOPs/VLOSEs?}
To answer the main research question, we developed the following sub-research questions:

\begin{itemize}
    \item[RQ1.] How can Article~34(1)(b) DSA be operationalized as a structured decision-tree-based toolkit for assessing negative effects on fundamental rights?
    \item[RQ2.] How can such a toolkit support the systematic identification of subsystems, stakeholders, and vulnerable positions in complex platform systems?
    \item[RQ3.] How can the toolkit structure the assessment and documentation of rights-relevant risks in a way that is reviewable and aligned with the DSA’s audit logic?
    \item[RQ4.] How can interdisciplinary expert feedback and persona-based evaluation reveal divergent assessment perspectives and support iterative toolkit refinement?
\end{itemize}

Our approach bridges legal interpretation and socio-technical evaluation, offering a methodological pathway for assessing structural harms, vulnerable positions~\cite{10.1145/3630106.3658915,10.1145/3630106.3658952} as central concerns in interdisciplinary research. We draw inspiration from existing toolkits for algorithmic fairness and equity~\cite{deng2022exploring, krafft2021action}, AI Act~\cite{hanif2024navigating}, and fundamental rights impact assessment~\cite{bogucka2024co, bogucka2025impact}.
as central concerns in interdisciplinary research. We draw inspiration from existing toolkits for algorithmic fairness and equity~\cite{deng2022exploring, krafft2021action}, AI Act~\cite{hanif2024navigating}, and fundamental rights impact assessment~\cite{bogucka2024co, bogucka2025impact}.
Based on work in computer science~\cite{10.1145/3442188.3445921}, including algorithmic auditing and accountability~\cite{10.1145/3173574.3174014, 10.1145/3593013.3594073, 10.1145/3630106.3658970}, human–computer interaction (HCI)~\cite{garcia2025default}, and legal compliance this toolkit aims to enable different users to identify~\cite{bogucka2025impact}, analyze, and assess 
the systemic risk of negative effects on fundamental rights.

\section{Background and Related Work}
\label{related_work}
Recent scholarship highlights how VLOPs/VLOSEs present distinct risks to democratic processes, vulnerable groups, and fundamental rights~\cite{wagner_mapping_2024, kubler_2021_2023, sanchez_playing_2025, 10.1145/3630106.3658970}. 
 Prior work has examined the scalability challenges of content moderation~\cite{gillespie_content_2020}, the limitations of transparency reporting~\cite{gillespie_custodians_2018}, and the potential of algorithmic audits for ensuring accountability~\cite{10.1145/3630106.3658970}. 
 While audit practices for AI systems are increasingly formalized~\cite{dobbe_hard_2021, marsden_platform_2022, raji2020closing}, the literature on designing processes, standards, or artifacts for supporting DSA compliance remains underdeveloped. In particular, methodological frameworks for assessing \textit{systemic risks} under Arts.~34 and 37 DSA are missing~\cite{sekwenz_doing_2025}. 
Related work on law-and-design card decks shows that practitioners benefit from tools that convert dense requirements into concise prompts that still preserve enough fidelity to support reflection and action~\cite{urquhart_right_2024}. This is especially relevant when the source material is difficult to operationalize directly in practice. At the same time, such translation always involves a tension between accessibility, granularity, and fidelity to the original requirements.

\paragraph{Systemic Risk Assessment under the DSA}
The DSA creates a novel auditing structure, primarily shaped through the introduction of two auditing circles – an internal (VLOP/VLOSE's systemic risk assessment) under Art.~34 DSA, and an external (auditor's independent audits) under Art.~37 DSA. 
Together, internal and external audits create a continuous accountability loop, but without shared standards their credibility depends on consistently translating audit tests into transparent, evidence-based methods for substantiating rights-relevant risks and mitigation~\cite{10.1145/3630106.3658970}.
Auditing under the DSA is resource-intensive and difficult to scale, which makes scalable approaches essential for meaningful oversight~\cite{sekwenz_cant_2025, 10.1145/3630106.3658970}.
These two (internal and external) auditing obligations are further regulated in the DR~\cite{noauthor_delegated_2023}, which provides details on methodologies, tests, evidence and risk for platforms and auditors~\cite{sekwenz_doing_2025}. Both audit obligations have annual reports as a final output (Arts.~34(1) and 37(4)~DSA, Art.~6~DR) and the two auditing instruments assess risk identification and mitigation, including on fundamental rights.
Operationalization of such methodologies in line with \emph{audit procedures}\footnote{Art. ~2(16)~DR which ``means any technique applied by the auditing organisation in the performance of the audit, including data collection, the choice and application of methodologies, such as tests and substantive analytical procedures, and any other action taken to collect and analyse information to collect audit evidence and formulate audit conclusions, not including the issuing of an audit opinion or of the audit report.” } 
and selected \emph{tests}\footnote{Art.~2(17)~DR defining ``[...] an audit methodology consisting in measurements, experiments or other checks, including algorithmic systems, through which the auditing organization assesses the audited provider’s compliance with the audited obligation or commitment.”} is a core component of the DR and necessary for meaningful compliance and assessment under the DSA.
Art.~34(1)(b)~DSA links systemic risk obligations to the Charter of Fundamental Rights by requiring providers to assess actual or foreseeable negative effects on the exercise of Charter-protected fundamental rights~\cite{RegulationEU20222022, european2000charter}. According to Recital~94 systemic risk assessments should follow a ``case-by-case basis" analyzing ``in-depth" (See also Recital~80~DSA). It leaves open how to conduct risk assessments, which rights to prioritize (intra-dependency of fundamental rights), and which indicators can credibly operationalize infringement in socio-technical systems (Recital~81~DSA). 
Fundamental rights might be directly liked to the scope of the DSA and named in the systemic risk provision or could be linked to it indirectly.\footnote{I.e. political rights – for example the right to freedom of thought, freedom of assembly and association, or the right to vote, or rights related to copyright and business - for example the right to property, freedom of arts and science, freedom to conduct a business.} 
For VLOPs, these rights imply obligations to protect users’ ability to communicate, access information, and participate without undue interference or harm, as operationalized in Art.~34(1)(b)~DSA (See Table~\ref{tab:eu-charter}), shifting EU digital governance from illegal content to the human rights impacts of e.g. design, recommender systems, and moderation at scale.

\paragraph{The Complexity of Assessing the Negative Effects on Fundamental Rights}
An additional complexity stems form the different test structures associated with fundamental rights, within their scope of application or case law based proportionality testing. 
Following Malgieri and Santos~\cite{malgieri_assessing_2025}, fundamental-rights risk assessment should distinguish \emph{violations}\footnote{In this text we also use the word ``infringement" for ``violation", see Section~\ref{Risk_Identification}.} from \emph{interferences}: while violations are typically treated as binary, interferences lie on a continuum across socio-technical interaction.
The severity of such interferences depends on contextual factors, including social norms (e.g. community standards of platforms), legislative interpretations (definition of illegal content), and how individuals or groups perceive the impact. As established by the Court of Justice of the European Union (CJEU) in several cases\footnote{Such as \textit{Digital Rights Ireland}, \textit{Google Spain}, \textit{Tele2 Sverige}, \textit{Ministerio Fiscal}, and \textit{La Quadrature du Net}, as well as by the ECtHR in \textit{Zakharov}.} the ``seriousness” of an interference is variable (different per human right) and measurable. This positions \emph{severity} as a central concept for proportionality analysis under Art.~52 Charter and for statutory obligations that require private entities to assess necessity and proportionality, such as Art.~35(7)(b) General Data Protection Regulation (GDPR)~\cite{noauthor_regulation_2016}. Accordingly, what must be assessed in practice is not the violation itself, often a legal conclusion, but the \emph{degree of interference} with fundamental rights. 
In line with Malgieri and Santos~\cite{malgieri_assessing_2025}, assessing risks to fundamental rights requires a clear conceptual alignment between infringements, interferences, violations, and harms. We adopt a normative lens that locates risks to fundamental rights in the (potential) \emph{infringement}, that is, if an \emph{interference} is not justified (See Section \ref{Risk_Identification} and proportionality testing steps 2a-4b). Interpreting these rights in socio-technical contexts demands interdisciplinary approaches. For example understanding freedom of expression entails studying content moderation; privacy relates to data infrastructures; and equality requires attention to bias studies. 
These systemic risk assessments shift fundamental-rights protection from a state–individual model to a multi-actor setting where VLOPs mediate rights at scale, regulators impose duties on private actors, and users experience interferences through platform design, ranking, and enforcement.
This conceptualization informs our approach to systemic risk assessments under the DSA, where understanding and quantifying the severity of interferences is essential for evaluating design choices, algorithmic effects, and mitigation measures.

\paragraph{Dimensions of Systemic Risks, Harms, Infringement, and Interference}

Within our design, the \emph{perception} dimension evaluates the normative seriousness of an infringement, considering the rules violated, its scope, duration, and reversibility. Second, the \emph{impact} dimension assesses how stakeholder (such as users, individuals, groups, and society) perceive the severity of the interference. 
To bridge risk-based, harm-based, and rights-based approaches, we apply a three-tiered assessment of severity (See Section \ref{Risk_Assessment} \emph{Severity}, \emph{Duration}, \emph{Reversibility}). 
The \emph{impact} dimension examines tangible impacts on affected persons or communities, including changes in well-being, financial loss, scale of impact, and the reversibility of those effects. Together, these tiers enable a granular and legally grounded assessment of severity that is essential for evaluating systemic risks under the DSA.
We differentiate these from \emph{harms} or adverse effects, which refer to measurable physical, psychological, economic, or social consequences for \emph{Risk Assessment} (See Section~\ref{Risk_Assessment}, \emph{Well-being Impact}, \emph{Reversibility Impact}, \emph{Financial Loss}, \emph{Reversibility of Financial Loss}). 
As illustrated in Figure~\ref{fig:wholeprocess}, the toolkit translates these dimensions into a structured decision-tree that guides the assessment from context and subsystem identification to the evaluation of objective seriousness, subjective and inter-subjective perception, and real-life consequences for affected persons and communities.

\paragraph{The Role of the Delegated Regulation on Conducting Audits in Toolkit Design}
An additional layer of regulatory complexity is the Delegated Regulation (DR)~\cite{noauthor_delegated_2023}. Our toolkit is explicitly designed to satisfy the DR’s core requirements for auditing systemic risks under Art.~34(1)(b)~DSA by translating them into a concrete \emph{audit procedure} (Art.~16~DR) with a clearly delimited \emph{audit scope} aligned with Art.~35~DSA (e.g., content moderation systems, recommender systems, advertising delivery, and related data practices)~\cite{noauthor_delegated_2023}. 
Methodologically, it combines \emph{quantitative and qualitative} components (Art.~10~DR), linking hypotheses (potential infringement of a human right) to assessable indicators (structured step-by-step-testing) while retaining contextual interpretation of negative effects of human rights (per subsystem, stakeholder, and human right), interference decision (outcome of the \emph{Risk Identification} Layer, see Section~\ref{Risk_Identification}), and their effects (See \emph{Risk Assessment} Layer~\ref{Risk_Assessment}).
To support auditable conclusions (e.g. audit evidence and quality criteria, Art.~11~DR), the toolkit operationalizes \emph{materiality threshold}~\footnote{‘materiality threshold’ means the threshold beyond which deviations or misstatements by the audited provider, individually or aggregated, would reasonably affect the audit findings, conclusions and opinions.} (Art.~2(12)~DR) by requiring explicit thresholds per risk category — and, where relevant, per affected users in vulnerable positions (Art.~10~DR)—so distributional impacts are made visible, not incidental. In line with the DR’s sequencing logic, identifying the groups impacted within situations that produce vulnerability precedes defining the audit scope per risk category (Art.~2~DR), e.g., specifying risks to LGBTQ+ politicians during electoral campaigns before selecting the subsystems and governance levers to examine~\cite{kubler_2021_2023}. Finally, the toolkit links sampling\footnote{See Art.~12~DR: ``Where audit evidence is based, partially or entirely, on a sample of data or information, the sample size and methodology for sampling shall be selected with a view to minimising the detection risk and without interference by the audited provider."} choices as a methodology in line with the DR to these scope decisions by requiring that samples are demonstrably adequate for the affected populations and the specific risk tested (Art.~12(12)(f)~DR).

\paragraph{Why use a toolkit? Risk Assessment Frameworks and Toolkits}
\label{sec:frameworks_toolkits}
According to other research decision tree-based toolkits are particularly suitable for non-lawyers due to its high level of comprehensibility (use of non-specialist language) and interpretability, especially in situations where a system (or subsystem) falls under several risk categories. ~\cite{hanif2024navigating, hupont2022landscape, krafft2021action}. With regard to the toolkit, other authors have pointed out that the user interface should be similar to comparable tools, should include context-sensitive alerts targeting common pitfalls and that guidelines and training materials can support practitioner~\cite{deng2022exploring}. Furthermore, it seems to be advicable to include checklists and questionnaire based interrogations. (Cf.~\cite{krafft2021action, deng2022exploring}
Toolkits bundle practical instruments (e.g., templates, worksheets, flowcharts, and questionnaires) to support the application of regulatory frameworks and to facilitate interdisciplinary cooperation~\cite{deng2022exploring}. Fairness toolkits\footnote{Examples of fairness-assessment toolkits include Fairlearn (\url{https://fairlearn.org/}), AIF360 (\url{https://ai-fairness-360.org/}), and Aequitas (\url{https://aequitas-home.readthedocs.io/}).} typically operationalise fairness through ready-to-use metrics to quantify undesirable bias in data or models, often alongside algorithmic techniques for measurement and mitigation~\cite{deng2022exploring}.

\paragraph{Toolkits in regulatory contexts}
Many EU governance frameworks (e.g., the GDPR, the AI Act, and the Cyber Resilience Act) require structured risk assessment processes. To enforce the right to the protection of personal data, the GDPR introduced in Art.~35~GDPR a formalised assessment mechanism that is frequently framed as a form of Fundamental Rights Impact Assessment (FRIA)~\cite{Guidelines32025, thomaidou2025navigating}. The AI Act~\cite{noauthor_regulation_2024_AIA} similarly adopts a
risk-based approach to ensure the development and deployment of trustworthy AI in line with fundamental rights and European values~\cite{union2021proposal, eu2021european, garcia2025default}. Finally, the EU Cyber Resilience Act~(CRA) introduces phased cybersecurity duties across the product lifecycle (including vulnerability handling, conformity assessment for certain products, and incident reporting)~\cite{noauthor_regulation_2024}. It anticipates overlap with the AI Act where products qualify as high-risk AI systems: CRA risk assessment should consider AI-specific vulnerabilities and, where relevant, fundamental-rights risks, while Art.~27~AI~Act requires a deployment-centred Fundamental Rights Impact Assessment. Together, these instruments establish a complementary logic: cyber-resilient products as a baseline, and context-specific FRIA at deployment to evidence (and mitigate) foreseeable rights-relevant
harms.
\paragraph{VLOPs/VLOSEs as Systems-of-Systems}
Large systems (such as VLOPs/VLOSEs) are typically not singular systems but \emph{systems-of-systems} (SoS): collections of multiple subsystems~\cite{sommerville2011software}, each posing distinct risks to stakeholders’ fundamental rights. With SoS complexity and non-deterministic behaviour, risk assessment approaches have shifted~\cite{sommerville2011software}. To manage complexity, it can be advisable to assess risks at the subsystem level first, noting that risks compound at higher levels when subsystem risks interact~\cite{freeman1997risk}. In parallel, design and engineering traditions have
moved from preventive to proactive, iterative approaches to risk management, including Boehm’s spiral model~\cite{boehm2000spiral}, Privacy by Design~\cite{cavoukian2009privacy}, Value-Sensitive Design~\cite{friedman2019value}, and Participatory Design~\cite{muller1993participatory}, each prescribing forms of assessment throughout system development.

\paragraph{Requirements of toolkits}
Deng et al.~(2022) identify three requirements for fairness toolkit design: interfaces should align with comparable tools and provide guidance and training materials; toolkits should include interactive, deliberation-oriented activities to stimulate critical reflection and interdisciplinary discussion; and toolkits should provide context-sensitive alerts and checklists to support task completion and to prompt critical conversations among stakeholders~\cite{deng2022exploring}.

Several recent toolkits operationalise legal requirements through decision-tree or template-based structures. Hanif et al.~\cite{hanif2024navigating} propose a decision-tree framework to classify AI systems into AI Act risk categories.\footnote{These risk categories are: unacceptable, high, limited and minimal.} Decision trees represent decisions paths, and outcomes: internal nodes encode decisions, branches encode resulting paths, and leaves encode final outcomes.Their interpretability can support classification by non-specialist practitioners~\cite{hanif2024navigating}, including cases where one system may implicate multiple categories~\cite{hanif2024navigating, hupont2022landscape}.
Bogucka et al.~(2024)~\cite{bogucka2024co} propose a standardised template to facilitate risk assessment aligned with the EU AI Act, NIST’s AI RMF, and ISO~42001. The toolkit follows a five-step process: (1) system context (intended use,purpose, capabilities/components, implementation context, and stakeholders); (2) potential risks (grouped across technical components, human interaction/experience, and systemic impacts on society, economy, and environment); (3)mitigation strategies; (4) anticipated benefits and positive impacts; and (5) reporting mechanisms and governance responsibilities~\cite{bogucka2024co}.
Similarly, the Algorithmic Equity Toolkit consists of three components~\cite{krafft2021action}. A flowchart (decision tree) supports scoping—identifying whether a technology constitutes an automated decision system. A questionnaire then structures interrogation of algorithmic harms and bias dimensions, and a system map/worksheet supports disentanglingintended purposes from misuse and clarifying how technical terms and components combine into an automated system~\cite{krafft2021action}.
Finally, Raol et al.~(2025)~\cite{rao2025riskrag} present RiskRAG, developed via (1) literature research and co-design to derive design requirements, (2) implementation based on those requirements, and (3) evaluation through user studies. Key requirements include real-world use cases and examples, recommendations of mitigation strategies, clear structuring of risks by impact/significance, and explicit linkage to use cases to support contextualized understanding of relevance and impact~\cite{rao2025riskrag}.

\paragraph{Why Vulnerability is a Critical Design Lens for the DSA}
Design decisions of these platforms directly impact individuals ~\cite{milton2023see, sala2024social, twenge2018increases}, societal norms and democratic processes ~\cite{narula2022virtual}. 
On an individual level the design of VLOPs/VLOSEs can introduce significant and varied risks to users,\footnote{See e.g. the definition for systemic risks Art.~34(1)(d)~DSA ``[...] serious negative consequences to the person’s physical and mental well-being."} such as reduced self-satisfaction and mental and physical well-being due to harmful algorithms~\cite{milton2023see, sala2024social} and even an increase in suicide rates~\cite{twenge2018increases}. 
The greater the power, ``... the greater responsibility to act fairly and protect the vulnerable party" including digital aspects of user's expression~\cite{malgieri2025scalable}.
However, many VLOPs’ business models commodify personal data~\cite{doi:10.1177/20539517211017308}, creating tensions with the EU Charter~\cite{custers2022priceless} and regimes such as the GDPR~\cite{GeneralDataProtection, malgieri_pricing_2018}.

The susceptibility to harm may render virtually \emph{any} user - or other system stakeholders - vulnerable, and manifest itself in the perceived need for more prolonged usage~\cite{course2021social}, time spent~\cite{liu2022time}, the feeling of restlessness~\cite{hussain2021associations}, negative feelings when hindered to use social media~\cite{truzoli2023social}, failure in usage restriction ~\cite{rixen2023loop}, sleep deprivation~\cite{alonzo2021interplay}, problems in user's social life~\cite{zheng2016excessive}, lost of usage feeling ~\cite{parry2021systematic}, or FOMO~\cite{oberst2017negative} 
are essential signals of user's vulnerability introduced by social media~\cite{helberger2022choice}.  
Besides, there are harms that occur in situations of malicious user behavior targeted towards other users, for example in doxing~\cite{snyder2017fifteen}, malicious reporting behavior ~\cite{zhao2019decade}, or trolling~\cite{zannettou2019let}. In this article however, we do not focus on such user-introduced harms, but rather on actual or foreseeable negative impacts on the exercise of fundamental rights, with a particular focus on system stakeholder in vulnerable positions.

\paragraph{Vulnerability as a Heightened Risk to Fundamental Rights}
Vulnerability is a core concept in understanding negative effects on fundamental rights and a measure explicitly asked for in tests under the DR for sampling methodologies in Art.~12(f)~DR.\footnote{``[T]he representation and appropriate analysis of concerns related to particular groups as appropriate, such as minors or vulnerable groups and minorities, in relation to the audited obligation or commitment.''} 
Building on Rebrean and Malgieri~\cite{rebrean2025vulnerability}, we define vulnerability as heightened fundamental-rights risk arising from power-imbalanced dependencies and shaped by personal and structural resilience—not an inherent deficit, but a cumulative condition and governance lens indicating where support or empowerment is needed. This aligns with HCI work treating vulnerability as a design and assessment lens~\cite{carli_vulnerability-oriented_2023, popova_vulnerability_2022}.
First, the notion of vulnerability as the \emph{higher risk to rights}, which expands the Commission’s notion of vulnerability as susceptibility to behavioral or decisional influence is broadly consistent with the DSA’s focus on systemic risks and harms, including protections for individuals in “vulnerable positions” (e.g., Rec.~3,~9).
Second, the component of power-imbalanced dependencies can be seen through potentially harmful dependency between users and services (e.g. negative effects on freedom of expression), but also through e.g. negative effects on privacy through the lack of data protection ~\cite{malgieri_digital_2025}. 
Third, the DSA relies on both structural and personal resilience, since platform mechanisms and users’ active engagement jointly enable processes such as user reporting~\cite{10.1145/3715275.3732036}.
Consequently, and aligning with other scholarship (e.g.,~\cite{helberger2022choice}, the experience of vulnerability in digital environments is contextual and situational, rather than forming on the basis of individual traits alone, as e.g. outlined in anti-discrimination law~\cite{noauthor_council_2000, noauthor_council_2000-1}. Additionally, the view of ''users in a vulnerable position'' connects with Value Sensitive Design’s conception of stakeholders as contextual roles rather than fixed entities~\cite{friedman2019value}, enabling analysis of dynamic user–system relations and showing how different roles can surface different risks. For Art.~34(1)(b)~DSA, these perspectives support treating “vulnerable positions” as relational and situational: assessments should identify platform-mediated dependencies and contextual factors that heighten rights risks, and justify mitigation in terms of both harm reduction and user protection, support, and empowerment.

\section{Methodology}
To develop our toolkit, we used a mixed-methods approach from design science to systematically examine and create artefacts for solving practical challenges~\cite{johannesson2021knowledge}. Following the first step used by Raol et al. 2025~\cite{rao2025riskrag} in developing their toolkit, we first conducted a literature review to identify the most important design requirements. Based on the requirements gathered, we developed a proposal for our toolkit. This proposal was then iteratively refined and improved based on the findings from a design thinking~\cite{brown2008design} workshop (See Section \ref{sec:Toolkit_Design}), in which the ‘rose, thorn and bud’ method~\cite{maRoseThornBud2021} was used with experts in this field and co-authors of this paper. 
These experts have backgrounds in social sciences, sociology, social law, and digital law, and have between 3 and 12 years of professional experience (See Table \ref{Workshop}). The feedback from the workshop served as a basis for discussion within the interdisciplinary team of authors as they further iterated on the inital design. Based on the feedback and discussions, the decision tree was restructured and refined to align it with legal, technical, and design requirements. Ultimately, the toolkit was implemented as a GitLab Page, \footnote{\url{https://dylwi.gitlab.io/resocial/}} running entirely in the browser with no back-end involved, thereby ensuring that no data is transmitted or stored outside the users device. The results can be exported as json or csv files for further analysis and reporting.

\begin{figure}
    \centering
    \includegraphics[width=0.8\linewidth]{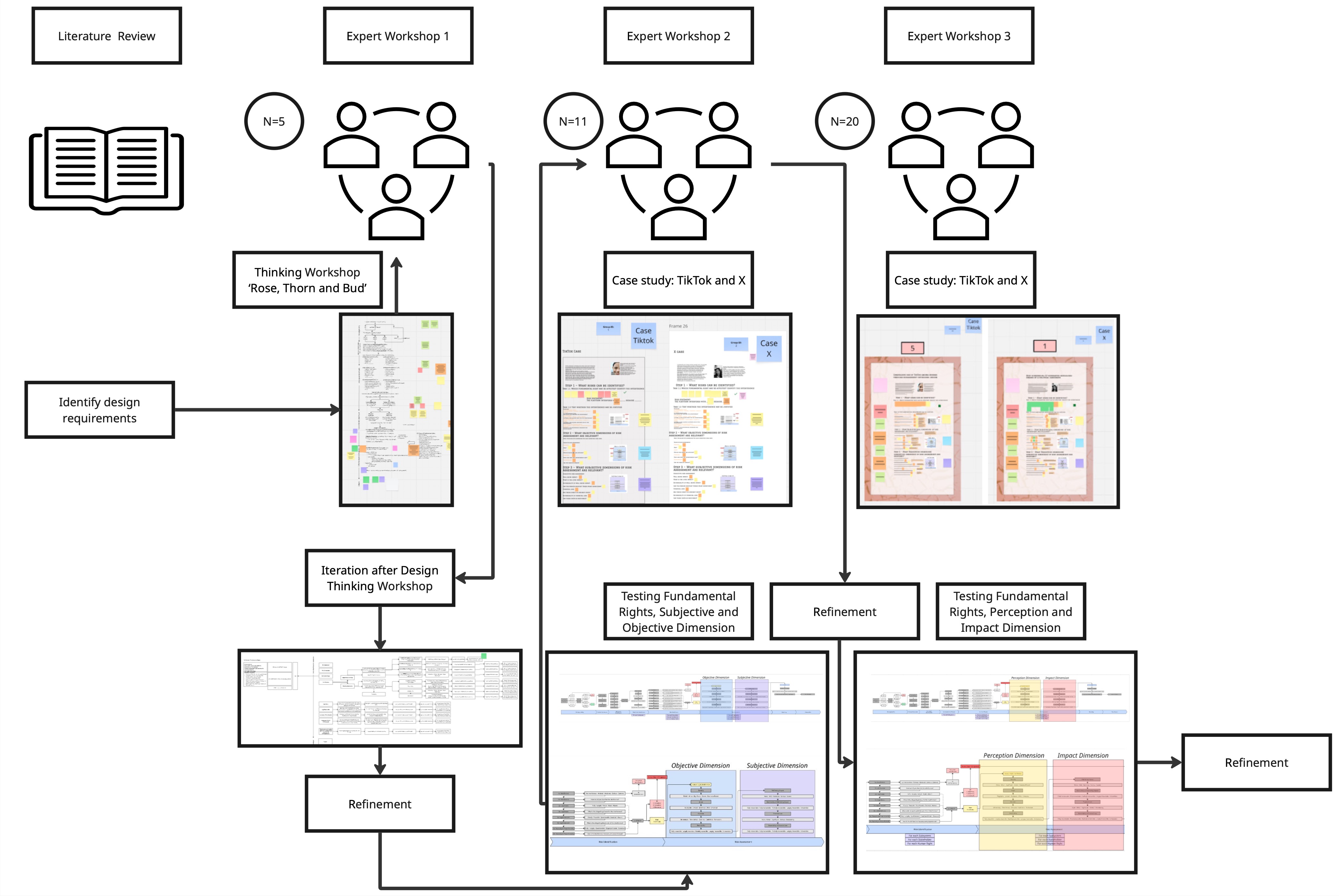}
    \caption{Methodological Overview}
    \Description{A methodological overview image.}
    \label{fig:workshop_info}
\end{figure}

\begin{table*}[t]
\caption{Overview of participants across the three toolkit evaluation workshops.}
\label{tab:workshop-composition}
\centering
\footnotesize
\setlength{\tabcolsep}{5pt}
\renewcommand{\arraystretch}{1.15}

\begin{tabularx}{\textwidth}{
@{}p{0.10\textwidth}
p{0.10\textwidth}
p{0.22\textwidth}
p{0.20\textwidth}
p{0.15\textwidth}
>{\raggedright\arraybackslash}X@{}
}
\toprule
\textbf{Workshop} &
\textbf{ID(s)} &
\textbf{Background / stakeholder group} &
\textbf{Position} &
\textbf{Experience} &
\textbf{\(n\)} \\
\midrule

Workshop 1 &
P1 &
Law &
Associate Professor &
10 years &
1 \\

&
P2 &
Sociology, socio-legal studies &
PhD Researcher &
3 years\textsuperscript{*} &
1 \\

&
P3 &
Social Science &
Full Professor &
12 years &
1 \\

&
P4 &
Law &
Research Assistant &
1 year &
1 \\

\addlinespace
\multicolumn{5}{@{}l}{\textit{Workshop 1 total}} &
\textbf{4} \\

\midrule

Workshop 2 &
P1--P11 &
Interdisciplinary experts &
-- &
-- &
\textbf{11} \\

\midrule

Workshop 3 &
P01--P11 &
Academic/research &
-- &
-- &
11 \\

&
P12 &
Public authority/regulator &
-- &
-- &
1 \\

&
P13--P16 &
Industry/private sector &
-- &
-- &
4 \\

&
P17--P20 &
Civil society &
-- &
-- &
4 \\

&
O01--O08 &
Organizers/facilitators &
-- &
-- &
8 \\

\addlinespace
\multicolumn{5}{@{}l}{\textit{Workshop 3 participant sample}} &
\textbf{20} \\

\multicolumn{5}{@{}l}{\textit{Workshop 3 total attendance incl.\ organizers/facilitators}} &
\textbf{28} \\

\midrule
\multicolumn{5}{@{}l}{\textbf{Total participant records across Workshops 1--3}} &
\textbf{35} \\

\multicolumn{5}{@{}l}{\textbf{Total attendance records incl.\ organizers/facilitators}} &
\textbf{43} \\

\bottomrule
\end{tabularx}

\vspace{2pt}
\raggedright
\footnotesize
\textsuperscript{*}Includes one year of policy-making experience.
\end{table*}

\paragraph{Ethical considerations.}
This study received approval from the relevant Human Research Ethics Committee (HREC) prior to data collection. The workshops involved low-risk professional feedback on the toolkit. Participants provided informed consent. Our analysis uses only anonymized contributions recorded on the workshop boards and summary notes prepared by the workshop organizers; no identifying information is reported. Data were handled in accordance with the approved procedures for consent, confidentiality, and data management.

\subsection{Case Law-Based Evaluation Use Cases}
\label{use_case_methodolgy}
To evaluate the toolkit, we use a deliberately extreme but plausible scenario involving non-consensual AI-generated sexualized images. The case is easy to understand, legally salient, and complex enough to activate multiple parts of the toolkit, including subsystem identification, stakeholder vulnerability, fundamental-rights interference, and mitigation.

The vignette, ``Undressed Without Consent,'' draws on \textit{Glawischnig-Piesczek v.\ Facebook Ireland} (C-18/18)\cite{noauthor_eva_2019},\footnote{We took the enforceability challenge of removal orders that can extend beyond identical uploads to ``equivalent” variants—directly relevant to re-uploads and near-duplicates. Furthermore, we use a case of ``variants” and how they can look like building on another leading case.}
LG Frankfurt am Main, 2-03 O 188/21 ~\cite{noauthor_burgerservice_nodate},\footnote{Which treats practical evasion tactics (e.g., altered layouts, added/removed text, typos, even minor pixel changes) as potentially still covered by a post-notice removal/locking obligation.}
and recent concerns around generative-AI ``undressing'' tools and repeated uploads of near-duplicate content ~\cite{field_grok_2026, milmo_grok_2026, noauthor_grok_2026, burgess_grok_nodate, gentleman_grok_2026, noauthor_groks_2026}. We use it as a stress test to examine whether the toolkit can structure complex and potentially divergent assessments rather than as a representative case of all DSA systemic risks.

The second vignette, ``Compulsive Use of TikTok Among Minors Through Engagement-Optimizing Design,'' examines systemic risks arising from platform design rather than from individual items of content. It is grounded in the European Commission's DSA proceedings concerning TikTok's potentially addictive design and protection of minors, including concerns about infinite scroll, autoplay, push notifications, and recommender systems \cite{europeancommission_tiktok_proceedings_2024,europeancommission_tiktok_addictive_2026}. We use the case to test whether the toolkit can capture interactions between platform design, recommender systems, user vulnerability, well-being, and mitigation measures.

\section{Findings}

\subsection{Toolkit Design Iterations}
\label{sec:Toolkit_Design}
\subsubsection{Initial Testing of the Toolkit}
\label{Workshop}
Four experts with interdisciplinary backgrounds participated in the first test workshop; their positions ranged from research assistant to full professor, and they had between 1 and 12 years of professional experience
(see Table in the Annex~\ref{tab:workshop-composition}).

\begin{figure}[ht]
    \centering
    \includegraphics[width=1\linewidth]{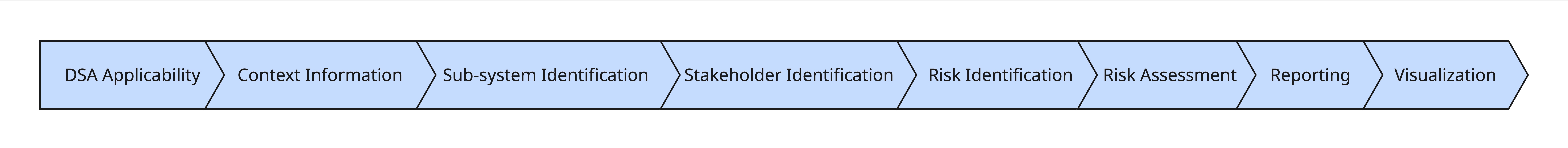}
    \caption{Overview of the Decision-Tree Toolkit Workflow}
    \Description{A workflow diagram showing the decision-tree toolkit as eight connected stages arranged from left to right in arrow-shaped boxes. The stages are: DSA applicability, context information, sub-system identification, stakeholder identification, risk identification, risk assessment, reporting, and visualization. The figure emphasizes the sequential progression of the toolkit from determining whether the DSA applies, through identifying relevant technical and social elements and assessing risks, to reporting and visualizing the results.}
    \label{overview_Structure_Toolkit}
\end{figure}

As a result of the ‘rose, thorn and bud’ method, we received 52 comments from four participants, of which 11.76\% were requests for examples to improve understandability, 17.65\% were references to limitations, 25.49\% were comments that the toolkit is good in its current form, and 45.10\% were suggestions for improvement. While the requested examples and most of the suggested improvements have been incorporated directly into the reworked version of the toolkit described in this section, the limitations are addressed in the ‘Limitations and Future Work’  section~\ref{limitation}.
Building on the step-by-step process described by Bogucka et al. (2024)~\cite{bogucka2024co}, the toolkit is structured as an eight-step process consisting of: 1) DSA applicability, 2) context information, 3) sub-system identification, 4) stakeholder identification, 5) risk identification, 6) risk assessment, 7) reporting, and 8) visualization. These steps are described in more detail in Section ~\ref{Workflow} and an overview is illustrated in Figure~\ref{overview_Structure_Toolkit}.

\subsubsection{Second Iteration of Toolkit Design}
The created persona descriptions were used, in conjunction with an export of the LimeSurvey toolkit’s structure, to feed the Anthropic ``Sonnet 4.6'' model,\footnote{\url{https://www.anthropic.com/claude/sonnet}} which then ran through the toolkit from the perspective of each individual persona. This generated CSV file as output, which could afterwards be further analyzed and used as the basis for visualizing the results in a deployment diagram using the open-source software PlantUML. The results are synthetic and cannot be considered empirical evidence; however, they can serve as an evaluation approach to simulate how real institutional actors—represented as personas—might respond to this toolkit, thereby providing insights for future design iteration.

As far as the interference score is concerned, there is a high degree of consensus among the personas regarding \emph{the rights of the child}, with this value rated at three or higher. The absence of an age verification feature for the AI editor was most frequently cited here as a key factor in the decision. 
There is near-unanimity among the personas regarding the interference score of \emph{human dignity} and \emph{respect for private and family life} (mean scores of 3.67 and 3.83 - between ``moderate'' and ``serious'' infringement), but their opinions diverge immediately when it comes to assessing proportionality. The European Commission, academic researcher, and NGO personas rate the legal basis and the legitimate aim as ``absent'', while the platform’s compliance team rates both as ``questionable''. 

The platform's compliance team persona tends to systematically downplay infringements on all fundamental rights—with the exception of \emph{freedom of expression and information}, with deltas ranging from -0.8 for \emph{human dignity} to -1.6 \emph{consumer protection}.
\emph{Consumer protection} and \emph{non-discrimination} are among the highly contested fundamental rights between the personas. While the independent auditor finds ``no infringements'', the nongovernmental organization classifies it as ``serious''. The differences between the various personas are even clearer when it comes to rating \emph{non-discrimination} infringement: The European Commission and DSC rate this as 3 (``moderate), non-governmental organizations as 5 (``extreme''), and academic researchers as 4 (``serious''), while the compliance team follows this tendency to downplay the issue and rates it as 2 (``minimal'').

With regard to the subjective and objective dimensions of risk assessment, there is a close correlation between the perceived severity of the human rights interference (subjective dimension) and the perceived impact on well-being (objective dimension). The more severe the effects, the greater the impact on well-being—or the greater the impairment of it. In the objective dimension, there is a clear difference in the assessment of financial loss. The platform’s compliance team rates ``financial loss'' as 4 (``serious''), while the DSC and the independent auditor rate it as 2 (``minor''). This could suggest that different personas interpreted the term ``financial loss'' differently, with the platform’s compliance team viewing financial losses as harm to the platform itself, while the others correctly understood it as financial harm to the users.

\begin{figure}[ht]
    \centering
    \includegraphics[width=1\linewidth]{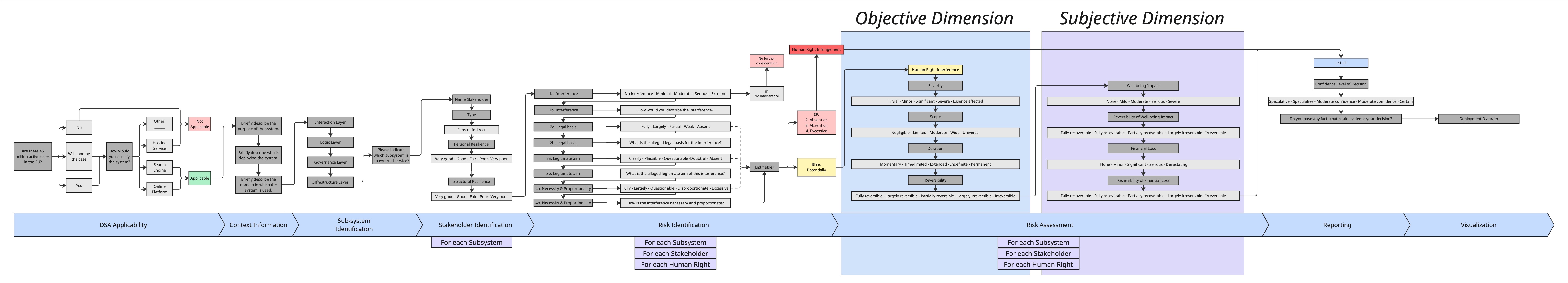}
    \caption{Overview of the decision-tree-based toolkit for assessing systemic risks to fundamental rights under Art.~34(1)(b) DSA, from DSA applicability and context definition to subsystem and stakeholder identification, risk identification, risk assessment, reporting, and visualization.}
    \Description{A full workflow diagram of the toolkit. The process runs from left to right through the stages DSA applicability, context information, subsystem identification, stakeholder identification, risk identification, risk assessment, reporting, and visualization. Each stage is represented by a blue banner, with detailed gray and colored boxes above indicating the questions, decision points, and outputs associated with that stage.}
    \label{fig:wholeprocess}
\end{figure}

\subsection{Final Workflow of the DSA-Decision Tree Toolkit}
\label{Workflow}

The workflow of the DSA-Decision tree toolkit operationalizes the assessment logic underlying Article 34(1)(b) DSA by structuring the evaluation into a sequence of connected steps, from applicability and context definition to subsystem mapping, stakeholder analysis, risk identification, risk assessment, reporting, and visualization. This stepwise structure is intended to support consistent, transparent, and auditable assessments of negative effects on fundamental rights in complex platform systems.

\subsubsection{DSA Applicability and Context Information}
In this step, it is first determined whether the system under investigation falls within the scope of the toolkit as proposed by Kraft (2021)~\cite{krafft2021action}.

In our case, this means checking whether the DSA applies or will be applicable in the foreseeable future. The DSA only applies to a) a specific type of system (VLOPs and VLOSEs) with b) 45 million active users in the EU, which can be implemented as a decision tree according to Art.~33.~DSA~\cite{RegulationEU20222022} Once the applicability of the toolkit has been determined, open-ended questions are used to gather contextual information about the system – purpose, provider, domain and core technical competencies (Cf.~\cite{bogucka2024co}). In line with the workshop findings, examples of how to define the system purpose are provided. These information can serve as a starting point for assessing the ``social, technical, cultural and other structural elements" that have led to the system use~\cite{rebrean2025vulnerability}, as necessary for understanding stakeholder in vulnerable position. Since the systems covered by the DSA are SOS, the next step is to examine the subsystems of which they consist.
We establish that the worked case is within scope of the toolkit by fixing DSA applicability assumptions (VLOP threshold) and the service classification as a very large online platform. 
As shown in Figure~\ref{fig:wholeprocess}, the toolkit operationalizes this initial screening as a simple decision tree that first considers the relevant user threshold and then the classification of the service under the DSA. In this way, the figure makes explicit that applicability is not assessed in the abstract, but through a structured sequence of questions that helps determine whether the system should be treated as within the scope of the toolkit.

\subsubsection{Context Information}
As shown in Figure~\ref{fig:wholeprocess}, the toolkit begins by gathering core contextual information about the system under assessment. This step captures three basic elements: the system’s purpose, the actor or organization deploying it, and the domain in which it is used. 
This step captures three basic elements: the system’s purpose, the actor or organization deploying it, and the domain in which it is used. Starting with these descriptors helps situate the assessment within the broader sociotechnical setting in which the system operates, rather than treating the system as an isolated technical artifact \cite{friedman2017survey,kallina2024stakeholder}. This is important because the interpretation of possible negative effects depends not only on technical functionality, but also on the context in which the system is developed, deployed, and encountered by different stakeholders \cite{kallina2024stakeholder}. The context-information step therefore provides the foundation for the subsequent identification of subsystems, stakeholders, and possible risks, while also supporting a more grounded interpretation of later findings.

\subsubsection{Subsystem Identification}
In order to address the complexity of \emph{System of Systems} in risk assessment, a series assessments must be carried out for each subsystem, for which these systems must be identified first~\cite{freeman1997risk}. The DSA names several subsystems that should be given special attention, including recommendation systems, content moderation systems, systems for enforcing terms of use, systems for delivering advertising, and their data-related systems.\footnote{See Art~34(2)~DSA
(a) the design of their recommender systems and any other relevant algorithmic system;
(b) their content moderation systems;
(c) the applicable terms and conditions and their enforcement;
(d) systems for selecting and presenting advertisements
(e) data related practices of the provider.}

For our initial decision tree framework, we differentiate between six layers that abstract from the DSA terminology, based on the feedback of our evaluation workshop this was reduced to four layers. The \emph{interaction layer} – also known as the front end – comprises systems such as web front ends or desktop applications, which are essential for the user experience, displaying information and user input. The \emph{logical layer} comprises recommendation, matching or indexing systems that work with the collected data and are essential for decision-making. The \emph{infrastructure layer} includes all systems that are required as a technical basis, such as monitoring and content delivery systems. Finally, the \emph{governance level} encompasses all systems responsible for enforcing rules, guidelines and standards, including systems such as content moderation and fraud detection systems. 

\subsubsection{Stakeholder Identification}
As shown in Figure~\ref{fig:wholeprocess}, the stakeholder-identification step operationalizes this broader understanding by documenting, for each subsystem, the relevant stakeholder, their type, whether they are directly or indirectly affected, and their levels of personal and structural resilience.
According to ISO 15288, stakeholders are all people who have ``a legitimate right, share,  claim, influence or interest in a system” (~\cite{iso2011ieee} p.~10). For an appropriate risk assessment, not only direct stakeholders, i.e., all persons who interact directly with a system, should be taken into account, but also indirect stakeholders who never or only rarely interact with a system but are nevertheless affected by it~\cite{iso2022ieee, friedman2019value}. Based on feedback from the workshop, clarification of the difference between direct and indirect stakeholders is provided, with examples.

Additionally, the view of ``users in a vulnerable position'' connects with Value Sensitive Design’s conception of stakeholders as contextual roles rather than fixed entities~\cite{friedman2019value}, we build upon a definition of vulnerability as heightened fundamental-rights risk arising from power-imbalanced dependencies~\cite{malgieri_digital_2025}. This conception is directly supported by Art.~13(1)(a)(f)~DR, requiring VLOPs provider to test the risk assumption with the most impacted groups.\footnote{It states that the size and methodology for sampling should be selected so that: ``the representation and appropriate analysis of concerns related to particular groups as appropriate, such as minors or vulnerable groups and minorities, in relation to the audited obligation or commitment."} Since the heightened fundamental-rights impact is shaped by personal and structural resilience, these must also be assessed for each stakeholder~\cite{malgieri_digital_2025}. As this is difficult without in-depth insights – assessment of the skills and knowledge of each individual stakeholder – we rely on assumptions about the stakeholders here.

\subsubsection{Risk Identification}
\label{Risk_Identification}

For this step we operationalize Article 52(1) of the Charter~\cite{european2000charter} by analyzing whether a subsystem of a platform  constitutes an interference with a fundamental rights protected by the Charter, and, if so, whether it  is provided for by law, pursues a legitimate aim, and  complies with the principle of proportionality~\cite{Charter_Explanations_2007_Art52} for a specific stakeholder.

The first step of this analysis consists of determining whether a subsystem of a platform interference with a right (see: Table~\ref{tab:eu-charter}) and limits its exercise, for example through practices such as content moderation.\footnote{I.e. as listed in Art.~17(1)(a)-(d)~DSA: restrictions of the visibility (removal of content, disabling access to content, or demoting content), suspension, termination or other restriction of monetary payments, of the service, or of the recipient of the service's account.}

Where an interference is classified (1a.) and described (1b.), the analysis must then determine, as a following step, whether that interference is ``provided by the law" within the meaning of article 52(1) of the Charter (2a.), by determining the normative basis on which the measure is adopted and the framework through which it is regulated (2b.), including obligations under EU law and the platform's terms and conditions as implemented pursuant to Article 14 of the DSA. In this regard, the applicable rules must be publicly accessible and formulated with sufficient clarity to ensure that their application is foreseeable~\cite{ECtHR_SundayTimes_v_UK_No1_1979}.

Third, it must be examined whether the measure can reasonably be connected to an objective of general interest recognised by the UE, or whether it pursues the protection of the rights and freedoms of others. In the context of risk assessment, this means that any measure relied as a mitigation step should be linked to this  objective and supported by a clear justification showing how it contributes to that aim - for example, by mitigation the dissemination of illegal content, safeguarding users’ rights, or protecting minors. If the reason relied upon cannot justify a restriction of the Charter's right, the measure will fail to satisfy Article~52(1) and will not be able to be treated as legitimate limitation of that right. 
Followed by assessing if there is a \emph{legitimate aim} (3a.) and what such a higher order objective might constitute (3b.).
The next step of this assessment is to verify that the measure complies with the principle of proportionality (4a. and 4b.). 
Under EU law, this criteria has been formulated as a structured assessment in which a measure must not exceed what is appropriate and necessary to attain their legitimate objectives~\cite{CJEU_Fedesa_C331_88_1990}. If more than one suitable option is available, the least restrictive alternative should therefore be preferred, and the resulting adverse effects must not be disproportionate in relation to the aim persued. The CJUE allowed to identify that the level of scrutiny is not uniform, but depends on both the right concerned and the gravity of the interference with, for example, a requirement of strict necessity and the presence of effective safeguards in the context of privacy and data protection~\cite{CJEU_DigitalRightsIreland_JoinedC293_12_C594_12_2014}. By contrast, in a situation where several rights or freedoms are engaged such as, for example, freedom of expression and economic interests, the assessments of proportionality may require explaining how the measure reconciles these interests~\cite{CJEU_Schmidberger_C112_00_2003}. 

\subsubsection{Risk Assessment}
\label{Risk_Assessment}
The tree structures the evaluation of rights-relevant risks \emph{for each subsystem}, \emph{for each stakeholder}, and \emph{for each human right} as illustrated in figure \ref{fig:wholeprocess}. Its starting point is whether a subsystem creates a \emph{human-rights interference} (See Section \ref{Risk_Identification} ). If so, the tree assesses that interference along two connected dimensions.
First, it evaluates the \emph{normative seriousness} of the interference itself through four criteria: \emph{severity} (from trivial to essence affected), \emph{scope} (from negligible to universal), \emph{duration} (from momentary to permanent), and \emph{reversibility} (from fully reversible to irreversible). This branch captures how serious the interference is as a matter of rights protection.
Second, it evaluates the \emph{practical consequences} of that interference. This includes \emph{well-being impact} and its reversibility, as well as \emph{financial loss} and its reversibility. These steps capture whether the interference produces tangible harms in the lives of affected stakeholders.

\subsubsection{Reporting}
This subsection explains how we translate decision-tree classifications into a documented risk assessment output that supports DSA systemic risk assessments (Art.~34 DSA) and is compatible with the Delegated Regulation’s audit requirements (Art.~37 DSA;~DR). 
For each outcome, the toolkit records the implicated subsystem(s), affected stakeholder roles (including vulnerable situations), relevant Charter rights, and the legal/proportionality rationale. We then separate \emph{risk signal severity} (magnitude of the rights-relevant risk) from \emph{assurance/confidence} (strength of the evidence and controls), and produce a short assessment summary. Severity is characterized via probability and impact (including differential distribution), evidence is mapped to claims and uncertainty is explicitly flagged where inputs are insufficient, and findings are weighed using audit-risk and materiality considerations to support reasonable (not absolute) assurance. These artefacts support audit scoping by making uncertainty explicit and by indicating where additional testing and provider evidence would be required under the DR.

We translate the subsystem and stakeholder mapping into a small set of reviewable risk statements by treating risks as subsystem-mediated pathways to interferences with Charter rights and assigning the \emph{objective} and \newtheorem{subjective} dimension. This should provide an overview of the assessment and collect reviewable discussion points. 
The resulting output of this step is a risk assessment record that preserves the decision-tree path and its legal rationale, records severity and other considerations of the assessor(Art.~13(3)~DR; Art.~11 DR), and explicitly states the level of assurance, audit risks, and materiality considerations (Art.~2(8)--(13)~DR; Art.~9~DR). This provides the empirical and procedural grounding required for interdisciplinary mitigation discussions and supports iterative, documented, and structured assessments across different roles over time ~\cite{bogucka2024co}.
Finally, reporting mechanisms and responsibilities for the governance of use are outline

\subsubsection{Visualization}
\label{visualization}
The deployment diagram makes the system architecture and its relationship to infringements on fundamental rights visible and traceable by showing exactly which subsytems are associated with which rights. Such a diagram can be a valuable communication tool—it promotes structured thinking during analysis and makes statements understandable to regulatory authorities, auditors, and other stakeholders. For example, it can help to discuss systemic risks location and identification among an interdisciplinary group, pinning down criteria of assessment and opening up a group discussion. This can also help regulators to track enforcement action due to risk categories, or enforcement severity. Additionally, such diagrams can help platforms and auditors to support their systemic risk assessment processes and independent audit evidencing. Future work however, should test these features with participants with different backgrounds to further refine the toolkit.

\section{Discussion}
To answer our core research question – \emph{How can a structured risk-assessment methodology, (operationalized as a decision-tree-based toolkit), improve the identification, assessment, and documentation of adverse effects on fundamental rights, (under Art. 34(1)(b) DSA) for VLOPs/VLOSEs?} – we interpret the results of our assessment method about the toolkit’s capacity to structure and document Art.~34(1)(b) fundamental-rights risk assessments for VLOPs/VLOSEs, while recognizing the limits of curated use case.

\paragraph{The Need for Constant, Iterative, Interdisciplinary Learning in DSA Research}
To answer RQ1 we operationalized the legal requirements form Article~34(1)(b) DSA as an initial assessment toolkit solution.
Our approach bridges legal interpretation and socio-technical evaluation, offering a methodological pathway for assessing structural harms, vulnerable positions~\cite{10.1145/3630106.3658915, 
10.1145/3630106.3658952} as central concerns in CHI-centered research.
The stakeholder-identification step further extends this system's operationalization through mapping, and by asking who is affected by each subsystem, whether they are directly or indirectly affected, and whether they are placed in a vulnerable position. This is important because vulnerability is not treated as a fixed attribute of a person or group, but as a situational and relational condition produced through platform-mediated dependencies, limited resilience, or asymmetric power relations. The toolkit therefore helps shift the assessment from a general list of affected users toward a more precise account of which stakeholders are exposed to which rights-relevant risks, through which subsystem, and under what conditions (RQ2).
Based on interdisciplinary work in computer science~\cite{10.1145/3442188.3445921}, including algorithmic auditing and accountability~\cite{10.1145/3173574.3174014, 10.1145/3593013.3594073, 10.1145/3630106.3658970}, human–computer interaction (HCI)~\cite{garcia2025default}, and legal compliance this toolkit enables users to identify~\cite{bogucka2025impact}, analyze, and assess 
the systemic risk of negative effects on fundamental rights. Therefore, the discussion should not end with initial design or deployment of such toolkits, but constantly iterate on (legal) design and meaningful implementation. Work bridging participatory design and end-user development emphasizes that systems must remain open to revision as contexts, needs, and tasks change over time~\cite{tetteroo_participatory_2024}.

\paragraph{Vulnerability as a Core Contribution to Human Rights Assessments}
The toolkit’s vulnerability lens is intended to avoid to reifying static understanding to `vulnerability' but rather understanding that we need flexible and situational concepts for assessing systemic risks under the DSA~\cite{helberger2022choice, malgieri_assessing_2025, rebrean2025vulnerability}. This framing helps make distributional impacts and materiality judgments explicit rather than implicit. This matters for Art.~34(1)(b) DSA in practice because the challenge is not merely listing Charter rights, but evidencing who is put in vulnerable positions by a given subsystem pathway, and with what severity and distribution. The findings suggest that the toolkit’s main value lies in decomposing VLOPs/VLOSEs as systems-of-systems surrounded by various stakeholders in diverging vulnerable positions and assessment realities. Instead of assessing a platform as a single object, the toolkit asks assessors to identify relevant interaction, logic, governance, and infrastructure layers (RQ2). This decomposition makes it possible to locate where rights-relevant risks arise, for example in an AI image-generation model, a recommender system, a reporting interface, a moderation workflow, or an advertising-delivery system.

\paragraph{Streamlining Systemic Risk Assessment Methodologies and Workflows}
Our decision tree-based approach is to our knowledge the first attempt to operationalize systemic risk assessment methodologies for Art.~34(1)(b)~DSA (See Section \ref{related_work}).Besides, RQ3 asked how the toolkit can structure the assessment and documentation of rights-relevant risks in a way that is reviewable and aligned with the DSA’s audit logic.  Each assessment records the subsystem, stakeholder, affected right, interference, justification analysis, severity, scope, duration, reversibility, and practical consequences.  By providing a structured approach building on established toolkits from interdisciplinary contexts we provide the first structured solution on how make concrete assessment steps instead of only contributing to abstract discussions on DSA auditing. 
The framework separates the identification of a rights-relevant risk from the strength of the evidence supporting that assessment. This is important because DSA audits require not only conclusions, but also evidence, uncertainty statements, and confidence levels. 
This solution based approach is needed since systemic risk assessments differ on methodological aspects from platform to platform. This lack of standardization can hinder cross platform comparisons of compliance thresholds and make the identification of best practices challenging for other VLOPs/VLOSEs, regulators, auditors, researchers, and civil society. Besides, how to conduct systemic risk assessments for negative effects for fundamental rights is not standardized yet. 
This means that the toolkit’s contribution is not only classificatory, but procedural. It creates a record that can be inspected by auditors, regulators, researchers, civil-society actors, or internal compliance teams. In this sense, the toolkit supports the DSA’s audit logic by making rights-relevant judgments traceable, contestable, and easier to compare across subsystems (See Section \ref{visualization}), cases, and audits over time.

\section{Limitations and Future Work}
\label{limitation}

Our contribution has several limitations. First, the operationalization of fundamental-rights impacts under Art.~34(1)(b) DSA remains legally and methodologically unsettled, and the toolkit may need revision as enforcement, audit practice, and case law develop. Second, the decision-tree architecture simplifies complex socio-technical dynamics and cannot replace legal, technical, or domain expertise. Decomposing VLOPs/VLOSEs into subsystems may also overlook risks emerging from interactions between systems, organizations, and people.
Third, the evaluation use cases simplify evolving real-world situations. In particular, the legal and technical context of generative-AI ``undressing'' tools is changing rapidly, including through the Digital Omnibus on AI \cite{eu_ai_omnibus_2026}. The use cases should therefore be understood as stress tests rather than static representations of the regulatory environment.
Finally, the workshop analysis relies on anonymized board contributions and organizer summary notes. Future work should evaluate the toolkit in real-world audits and explore agentic AI for scaling subsystem mapping, evidence collection, and documentation \cite{sekwenz_cant_2025}. Such approaches should preserve human oversight, traceability, and a vulnerability lens, while supporting recurring and longitudinal assessment across large systems-of-systems.

\section{Conclusion}
This paper introduced a decision-tree-based toolkit for assessing systemic risks involving negative effects on fundamental rights under Article~34(1)(b) DSA. By translating legal and audit-oriented requirements into a structured workflow—including context information, subsystem and stakeholder identification, risk identification, risk assessment, reporting, and visualization—the toolkit offers a practical method for making complex platform assessments more traceable, transparent, and reviewable. In doing so, it contributes a first step toward operationalizing fundamental-rights risk assessment for VLOPs and VLOSEs in a way that is sensitive to socio-technical complexity, vulnerability, and the evidentiary demands of DSA compliance. At the same time, the toolkit is intended as an extensible methodological foundation rather than a final solution. Future work should therefore validate its use in practitioner settings, refine its application across different platform architectures and risk domains, and further examine how such structured approaches can support more consistent and auditable systemic-risk governance under the DSA.

\bibliographystyle{ACM-Reference-Format}
\bibliography{sample-base}
\section{Generative AI Usage Statement}
In line with the ACM policy, we used generative AI tools (including large language models) only for limited assistance with writing mechanics and formatting. Specifically, we used these tools to suggest grammar and fluency improvements to author-written text  and to assist with LaTeX formatting tasks. We did not use generative AI to generate substantive content, arguments, results, analyses, or interpretations. Generative AI was used to create personas and utilize them to populate the toolkit for generating synthetic data. All suggested edits were reviewed, selectively adopted, and verified by the authors, who remain fully responsible for the final manuscript.

\appendix
\section{Appendix}

\begin{table}[ht]
\centering
\caption{Relevant Articles of the EU Charter of Fundamental Rights based on the Fundamental Rights named in Art 34(1)(b)DSA}
\label{tab:eu-charter}
\footnotesize
\begin{tabular}{p{1.2cm} p{4.5cm} p{6.5cm}}
\toprule
\textbf{Article} & \textbf{Title} & \textbf{Short Description} \\
\midrule
\textbf{Art. 1} & Human Dignity & Human dignity is inviolable. It must be respected and protected. \\

\textbf{Art. 7} & Respect for Private and Family Life & Everyone has the right to respect for their private and family life, home and communications. \\

\textbf{Art. 8} & Protection of Personal Data & Everyone has the right to the protection of personal data concerning them. Such data must be processed fairly, for specified purposes, and based on consent or other legitimate basis laid down by law. \\

\textbf{Art. 11} & Freedom of Expression and Information & Everyone has the right to freedom of expression. This right includes freedom to hold opinions and to receive and impart information and ideas without interference by public authority. Media freedom and pluralism shall be respected. \\

\textbf{Art. 21} & Non-discrimination & Any discrimination based on grounds such as sex, race, colour, ethnic or social origin, genetic features, language, religion or belief, political or other opinion, membership of a national minority, property, birth, disability, age or sexual orientation shall be prohibited. \\

\textbf{Art. 24} & The Rights of the Child & Children shall have the right to such protection and care as is necessary for their well-being. They may express their views freely, and such views shall be considered in matters which concern them. \\

\textbf{Art. 38} & Consumer Protection & Union policies shall ensure a high level of consumer protection. \\
\bottomrule
\end{tabular}
\end{table}

\section{Use Case Description}

\begin{table*}[ht]
\centering
\caption{Use cases and personas used in the toolkit evaluation.}
\label{tab:use-cases}
\footnotesize
\begin{tabularx}{\textwidth}{p{0.22\textwidth} X X}
\toprule
\textbf{Use case} & \textbf{Case description} & \textbf{Persona} \\
\midrule

\textbf{Compulsive Use of TikTok Among Minors Through Engagement-Optimizing Design}
&
TikTok operates a personalized short-video feed optimized for engagement. The system uses infinite scroll, autoplay, algorithmic recommendations, notifications, streak-like interaction cues, and emotionally salient content to keep users active. Parents, teachers, and youth mental-health organizations report that minors are spending excessive time on the platform, including late at night. Some experience sleep disruption, reduced concentration, anxiety, body-image concerns, and difficulty disengaging. TikTok argues that it provides screen-time tools, parental controls, and ``take a break'' reminders. Critics argue that these protections are optional and structurally weaker than the platform's engagement-optimizing design.
&
\textbf{Sarah: Adolescent user exposed to engagement-optimizing platform design.}

Sarah is a minor TikTok user whose capacity to disengage is reduced by age, developing self-regulation, and possible mental-health vulnerabilities such as anxiety, loneliness, or body-image concerns. She uses TikTok not only for entertainment, but also for social belonging, peer recognition, and emotional regulation.
\\

\addlinespace

\textbf{Non-Consensual AI-Generated Sexualized Images of a Political Candidate}
&
X integrates an AI image-editing chatbot into its platform. Users can upload or reply to public photos and prompt the system to generate sexualized or ``undressed'' versions of the person depicted. During an Austrian election campaign, a young female political candidate becomes the target of repeated non-consensual AI-generated nude images. The images are reposted as screenshots, cropped versions, memes, and near-duplicates. The candidate reports the content, but removals are slow and copies continue to circulate. X argues that it has restricted the tool to paying users and improved reporting options. Civil society groups argue that this does not address the core risk: the platform makes non-consensual sexualized image generation technically permissible and socially scalable.
&
\textbf{Elena: Young female political candidate targeted through AI-generated sexualized abuse.}

Elena is a young female political candidate who uses X to communicate with voters, journalists, party members, and the wider public during an election campaign. Because of her public visibility and political role, she is exposed to intensified scrutiny, coordinated harassment, and gendered forms of abuse.
\\

\bottomrule
\end{tabularx}
\end{table*}

\end{document}